\documentclass[sigconf]{acmart}
\usepackage{booktabs} 

\usepackage{graphicx}
\usepackage{graphics}
\usepackage{algorithm}
\usepackage{algorithmicx}
\usepackage{algpseudocode}
\usepackage{caption}
\usepackage{subfigure}
\usepackage{amsmath}
\usepackage{amstext}
\usepackage{amssymb}
\usepackage{framed}
\usepackage{balance}
\usepackage{multirow}
\newtheorem{problem}{Problem}

\newcommand{\RN}[1]{\textup{\uppercase\expandafter{\romannumeral#1}}}
\newcommand{\mat}[1]{{\bf #1}}   
\graphicspath{{./figures/}}

\begin{document}
\title{Attributed Network Embedding for Learning \\
in a Dynamic Environment}

\author{Jundong Li}
\affiliation{%
  \institution{Arizona State University}
}
\email{jundongl@asu.edu}

\author{Harsh Dani}
\affiliation{%
  \institution{Arizona State University}
  }
\email{hdani@asu.edu}

\author{Xia Hu}
\affiliation{%
  \institution{Texas A\&M University}
  }
\email{hu@cse.tamu.edu}

\author{Jiliang Tang}
\affiliation{%
  \institution{Michigan State University}
}
\email{tangjili@msu.edu}

\author{Yi Chang}
\affiliation{%
  \institution{Huawei Research America}
  }
\email{yichang@acm.org}

\author{Huan Liu}
\affiliation{%
  \institution{Arizona State University}
  }
\email{huan.liu@asu.edu}

\acmConference{CIKM'17}{}{November 6--10, 2017, Singapore.}
\acmPrice{15.00}
\acmDOI{https://doi.org/10.1145/3132847.3132919}
\acmISBN{ISBN 978-1-4503-4918-5/17/11}


\renewcommand{\shortauthors}{J. Li et al.}

\begin{abstract}
Network embedding leverages the node proximity manifested to learn a low-dimensional node vector representation for each node in the network. The learned embeddings could advance various learning tasks such as node classification, network clustering, and link prediction. Most, if not all, of the existing works, are overwhelmingly performed in the context of plain and static networks. Nonetheless, in reality, network structure often evolves over time with addition/deletion of links and nodes. Also, a vast majority of real-world networks are associated with a rich set of node attributes, and their attribute values are also naturally changing, with the emerging of new content patterns and the fading of old content patterns. These changing characteristics motivate us to seek an effective embedding representation to capture network and attribute evolving patterns, which is of fundamental importance for learning in a dynamic environment. To our best knowledge, we are the first to tackle this problem with the following two challenges: (1) the inherently correlated network and node attributes could be noisy and incomplete, it necessitates a robust consensus representation to capture their individual properties and correlations; (2) the embedding learning needs to be performed in an online fashion to adapt to the changes accordingly. In this paper, we tackle this problem by proposing a novel dynamic attributed network embedding framework - DANE. In particular, DANE first provides an offline method for a consensus embedding and then leverages matrix perturbation theory to maintain the freshness of the end embedding results in an online manner. We perform extensive experiments on both synthetic and real attributed networks to corroborate the effectiveness and efficiency of the proposed framework.
\end{abstract}


\maketitle
\section{Introduction}
Attributed networks are ubiquitous in myriad of high impact domains, ranging from social media networks, academic networks, to protein-protein interaction networks. In contrast to conventional plain networks where only pairwise node dependencies are observed, nodes in attributed networks are often affiliated with a rich set of attributes. For example, in scientific collaboration networks, researchers collaborate and are distinct from others by their unique research interests; in social networks, users interact and communicate with others and also post personalized content. It has been widely studied and received that there exhibits a strong correlation among the attributes of linked nodes~\cite{shalizi2011homophily,pfeiffer2014attributed}. The root cause of the correlations can be attributed to social influence and homophily effect in social science theories~\cite{marsden1993network,mcpherson2001birds}. Also, many real-world applications, such as node classification, community detection, topic modeling and anomaly detection~\cite{jian2017toward,yang2009combining,li2017toward,li2017residual,he2017modeling}, have shown significant improvements by modeling such correlations.

Network embedding~\cite{chen2007directed,perozzi2014deepwalk,tang2015line,cao2015grarep,chang2015heterogeneous,grover2016node2vec,huang2017label,huang2017accelerated,qu2017attention} has attracted a surge of research attention in recent years. The basic idea is to preserve the node proximity in the embedded Euclidean space, based on which the performance of various network mining tasks such as node classification~\cite{aggarwal2011node,bhagat2011node}, community detection~\cite{tang2008community,yang2009combining}, and link prediction~\cite{liben2007link,wang2011human,barbieri2014follow} can be enhanced. However, a vast majority of existing work are predominately designed for plain networks. They inevitably ignore the node attributes that could be potentially complementary in learning better embedding representations, especially when the network suffers from high sparsity. In addition, a fundamental assumption behind existing network embedding methods is that networks are static and given a prior. Nonetheless, most real-world networks are intrinsically dynamic with addition/deletion of edges and nodes; examples include co-author relations between scholars in an academic network and friendships among users in a social network. Meanwhile, similar as network structure, node attributes also change naturally such that new content patterns may emerge and outdated content patterns will fade. For example, humanitarian and disaster relief related topics become popular on social media sites after the earthquakes as users continuously post related content. Consequently, other topics may receive less public interests. In this paper, we refer this kind of networks with both network and node attribute value changes as \emph{dynamic attributed networks}.

Despite the widespread of dynamic attributed networks in real-world applications, the study in analyzing and mining these networks are rather limited. One natural question to ask is when attributed networks evolve, how to correct and adjust the staleness of the end embedding results for network analysis, which will shed light on the understanding of their evolving nature. However, dynamic attributed network embedding remains as a daunting task, mainly because of the following reasons: (1) Even though network topology and node attributes are two distinct data representations, they are inherently correlated. In addition, the raw data representations could be noisy and even incomplete, individually. Hence, it is of paramount importance to seek a noise-resilient consensus embedding to capture their individual properties and correlations; (2) Applying offline embedding methods from scratch at each time step is time-consuming and cannot seize the emerging patterns timely. It necessitates the design of an efficient online algorithm that can give embedding representations promptly.

\begin{figure*}[!htbp]
\centering
\includegraphics[width=0.9\textwidth]{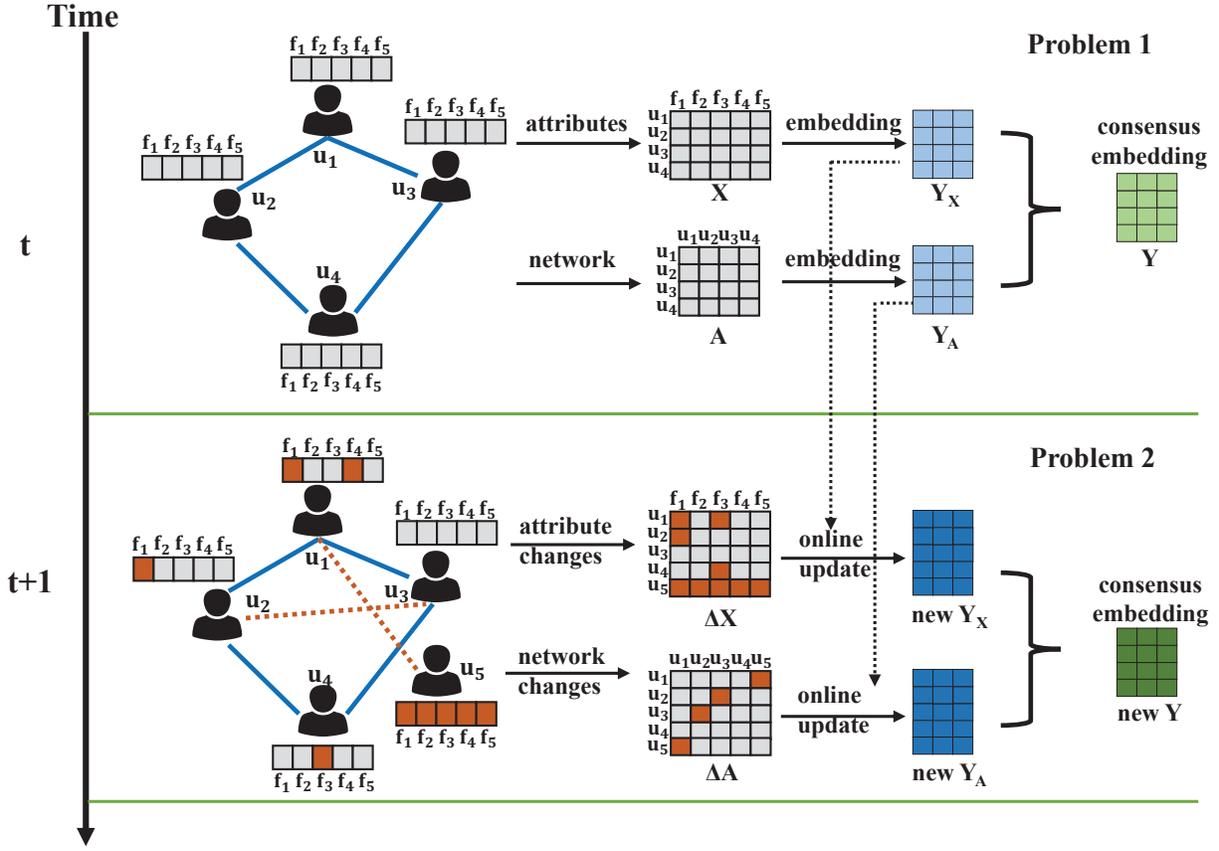}
\caption{An illustration of the proposed dynamic attributed network embedding framework - DANE. At time step $t$, DANE performs spectral embedding on network structure $\mat{A}$ and node attributes $\mat{X}$, and obtain two embeddings $\mat{Y}_{\mat{A}}$ and $\mat{Y}_{\mat{X}}$. Afterwards, DANE maximizes their correlation for a consensus embedding representation $\mat{Y}$. At the following time step $t+1$, the network is characterized by both topology structure and attribute value changes $\mat{\Delta}\mat{A}$ and $\mat{\Delta}\mat{X}$ (the changes are highlighted in orange). DANE leverages matrix perturbation theory to update $\mat{Y}_{\mat{A}}$ and $\mat{Y}_{\mat{X}}$, and give the new consensus embedding $\mat{Y}$.}
\label{fig:framework}
\end{figure*}

To tackle the aforementioned challenges, we propose a novel embedding framework for dynamic attributed networks. The main contributions can be summarized as follows:
\begin{itemize}
\item \textbf{\emph{Problem Formulations}}: we formally define the problem of dynamic attributed network embedding. The key idea is to initiate an offline model at the very beginning, based on which an online model is presented to maintain the freshness of the end attributed network embedding results.
\item \textbf{\emph{Algorithms and Analysis}}: we propose a novel framework - DANE for dynamic attributed network embedding. Specifically, we introduce an offline embedding method as a base model to preserve node proximity in terms of both network structure and node attributes for a consensus embedding representation in a robust way. Then to timely obtain an updated embedding representation when both network structure and attributes drift, we present an online model to update the consensus embedding with matrix perturbation theory. We also theoretically analyze its time complexity and show its superiority over offline methods.
\item \textbf{\emph{Evaluations}}: we perform extensive experiments on both synthetic and real-world attributed networks to corroborate the efficacy in terms of two network mining tasks (both unsupervised and supervised). Also, we show its efficiency by comparing it with other baseline methods and its offline counterpart. In particular, our experimental results show that the proposed method outperforms the best competitors in terms of both clustering and classification performance. Most importantly, it is much faster than competitive offline embedding methods.
\end{itemize}

The rest of this paper is organized as follows. The problem statement of dynamic attributed network embedding is introduced in Section 2. Section 3 presents the proposed framework DANE with analysis. Experiments on synthetic and real datasets are presented in Section 4 with discussions. Section 5 briefly reviews related work. Finally, Section 6 concludes the paper and visions the future work.

\section{Problem Definition}
We first summarize some notations used in this paper. Following the commonly used notations, we use bold uppercase characters for matrices (e.g., $\mat{A}$), bold lowercase characters for vectors (e.g., $\mat{a}$), normal lowercase characters for scalars (e.g., $a$). Also we represent the $i$-th row of matrix $\mat{A}$ as $\mat{A}(i,:)$, the $j$-th column as $\mat{A}(:,j)$, the ($i,j$)-th entry as $\mat{A}(i,j)$, transpose of $\mat{A}$ as $\mat{A}'$, trace of $\mat{A}$ as $tr(\mat{A})$ if it is a square matrix. $\mat{1}$ denotes a vector whose elements are all 1 and $\mat{I}$ denotes the identity matrix. The main symbols used throughout this paper are listed in Table~\ref{table:symbols}.
\begin{table}[!htbp]
\small
\begin{tabular}{|c|c|} \hline
Notations& Definitions or Descriptions \\ \hline \hline
$\mathcal{G}^{(t)}$ & attributed network at time step $t$ \\ \hline
$\mathcal{G}^{(t+1)}$ & attributed network at time step $t+1$ \\ \hline
$\mat{A}^{(t)}$ & adjacency matrix for the network structure in $\mathcal{G}^{(t)}$\\ \hline
$\mat{X}^{(t)}$ & attribute information in $\mathcal{G}^{(t)}$\\ \hline
$\mat{A}^{(t+1)}$ & adjacency matrix for the network structure in $\mathcal{G}^{(t+1)}$\\ \hline
$\mat{X}^{(t+1)}$ & attribute information in $\mathcal{G}^{(t+1)}$\\ \hline
$\Delta\mat{A}$ & change of adjacency matrix between time steps $t$ and $t+1$ \\ \hline
$\Delta\mat{X}$ & change of attribute values between time steps $t$ and $t+1$ \\ \hline
$n$ & number of instances (nodes) in $\mathcal{G}^{(t)}$ \\ \hline
$d$ & number of attributes in $\mathcal{G}^{(t)}$\\ \hline
$k$ & embedding dimension for network structure or attributes \\ \hline
$l$ & final consensus embedding dimension \\ \hline
\end{tabular}
\caption{Symbols.}
\vspace{-1\baselineskip}
\label{table:symbols}
\end{table}

Let $\mathcal{U}^{(t)}=\{u_{1},u_{2},...,u_{n}\}$ denote a set of $n$ nodes in the attributed network $\mathcal{G}^{(t)}$ at time step $t$. We use the adjacency matrix $\mat{A}^{(t)}\in \mathbb{R}^{n\times n}$ to represent the network structure of $\mathcal{U}^{(t)}$. In addition, we assume that nodes are affiliated with $d$-dimensional attributes $\mathcal{F}=\{f_{1},f_{2},...,f_{d}\}$ and $\mat{X}^{(t)}\in \mathbb{R}^{n\times d}$ denotes the node attributes. At the following time step, the attributed network is characterized with both topology and content drift such that new/old edges and nodes may be included/deleted, and node attribute values could also change. We use $\Delta\mat{A}$ and $\Delta\mat{X}$ to denote the network and attribute value changes between two consecutive time step $t$ and time step $t+1$, respectively. Following the settings of~\cite{tong2008colibri}, and for the ease of presentation, we consider the number of nodes is constant over time, but our method can be naturally extended to deal with node addition/deletion scenarios. As mentioned earlier, node attributes are complementary in mitigating the network sparsity for better embedding representations. Nonetheless, employing offline embedding methods repeatedly in a dynamic environment is time-consuming and cannot seize the emerging/fading patterns promptly, especially when the networks are of large-scale. Therefore, developing an efficient online embedding algorithm upon an offline model is fundamentally important for dynamic network analysis, and also could benefit many real-wold applications. Formally, we define the dynamic attributed embedding problem as two sub-problems as follows. The work flow of the proposed framework DANE is shown in Figure~\ref{fig:framework}.

\begin{problem}{The offline model of DANE at time step $t$: given network topology $\mat{A}^{(t)}$ and node attributes $\mat{X}^{(t)}$; output attributed network embedding $\mat{Y}^{(t)}$ for all nodes.}
\label{problem:problem1}
\end{problem}

\begin{problem}{The online model of DANE at time step $t+1$: given network topology $\mat{A}^{(t+1)}$ and node attributes $\mat{X}^{(t+1)}$, and intermediate embedding results at time step $t$; output attributed network embedding $\mat{Y}^{(t+1)}$ for all nodes.}
\label{problem:problem2}
\end{problem}

\section{The Proposed Framework - DANE}
In this section, we first present an offline model that works in a static setting to tackle the Problem~\ref{problem:problem1} in finding a consensus embedding representation. Then to tackle the Problem~\ref{problem:problem2}, we introduce an online model that provides a fast solution to update the consensus embedding on the fly. At the end, we analyze the computational complexity of the online model and show its superiority over the offline model.
\subsection{DANE: Offline Model}
Network topology and node attributes in attributed networks are presented in different representations. Typically, either of these two representations could be \emph{incomplete} and \emph{noisy}, presenting great challenges to embedding representation learning. For example, social networks are very sparse as a large amount of users only have a limited number of links~\cite{adamic2000power}. Thus, network embedding could be jeopardized as links are inadequate to provide enough node proximity information. Fortunately, rich node attributes are readily available and could be potentially helpful to mitigate the network sparsity in finding better embeddings. Hence, it is more desired to make these two representations compensate each other for a consensus embedding. However, as mentioned earlier, both representations could be noisy and the existence of noise could degenerate the learning of consensus embedding. Hence, it motivates us to reduce the noise of these two raw data representations before learning consensus embedding.

Let $\mat{A}^{(t)}\in\mathbb{R}^{n\times n}$ be the adjacency matrix of the attributed network at time step $t$ and $\mat{D}_{\mat{A}}$ be the diagonal matrix with $\mat{D}_{\mat{A}}^{(t)}(i,i)=\sum_{j=1}^{n}\mat{A}^{(t)}(i,j)$, then $\mat{L}_{\mat{A}}^{(t)}=\mat{D}_{\mat{A}}^{(t)}-\mat{A}^{(t)}$ is a Laplacian matrix. According to spectral theory~\cite{belkin2001laplacian,von2007tutorial}, by mapping each node in the network to a $k$-dimensional embedded space, i.e., $\mat{y}_{i}\in \mathbb{R}^{k}$ ($k\ll n$), the noise in the network can be substantially reduced. A rational choice of the embedding $\mat{Y}_{\mat{A}}^{(t)}=[\mat{y}_{1},\mat{y}_{2},...,\mat{y}_{n}]'\in\mathbb{R}^{n\times k}$ is to minimize the loss $\frac{1}{2}\sum_{i,j}\mat{A}^{(t)}(i,j)||\mat{y}_{i}-\mat{y}_{j}||_{2}^{2}$. It ensures that connected nodes are close to each other in the embedded space. In this case, the problem boils down to solving the following generalized eigen-problem $\mat{L}_{\mat{A}}^{(t)}\mat{a}=\lambda\mat{D}_{\mat{A}}^{(t)}\mat{a}$. Let $\mat{a}_{1},\mat{a}_{2},...,\mat{a}_{n}$ be the eigenvectors of the corresponding eigenvalues $0=\lambda_{1}\leq\lambda_{2}\leq...\leq\lambda_{n}$. It is easy to verify that $\mat{1}$ is the only eigenvector for the eigenvalue $\lambda_{1}=0$. Then the $k$-dimensional embedding $\mat{Y}_{\mat{A}}^{(t)}\in\mathbb{R}^{n\times k}$ of the network structure is given by the top-$k$ eigenvectors starting from $\mat{a}_{2}$, i.e., $\mat{Y}_{\mat{A}}^{(t)}=[\mat{a}_{2},...,\mat{a}_{k},\mat{a}_{k+1}]$. For the ease of presentation, in the following parts of the paper, we refer these $k$ eigenvectors and their eigenvalues as the top-$k$ eigenvectors and eigenvalues, respectively. Akin to the network structure, noise in the node attributes can be reduced in a similar fashion. Specifically, we first normalize attributes of each node and obtain the cosine similarity matrix $\mat{W}^{(t)}$. Afterwards, we obtain the top-$k$ eigenvectors $\mat{Y}_{\mat{X}}^{(t)}=[\mat{b}_{2},...,\mat{b}_{k+1}]$ of the generalized eigen-problem corresponding to $\mat{W}^{(t)}$.

The noisy data problem is resolved by finding two intermediate embeddings $\mat{Y}_{\mat{A}}^{(t)}$ and $\mat{Y}_{\mat{X}}^{(t)}$. We now take advantage of them to seek a consensus embedding. However, since they are obtained individually, these two embeddings may not be compatible and in the worst case, they may be independent of each other. To capture their interdependency and to make them compensate each other, we propose to maximize their correlations (or equivalently minimize their disagreements)~\cite{hardoon2004canonical}. In particular, we seek two projection vectors $\mat{p}_{\mat{A}}^{(t)}$ and $\mat{p}_{\mat{X}}^{(t)}$ such that the correlation of $\mat{Y}_{\mat{A}}^{(t)}$ and $\mat{Y}_{\mat{X}}^{(t)}$ is maximized after projection. It is equivalent to solving the following optimization problem:
\begin{equation}
\begin{split}
&\max_{\mat{p}_{\mat{A}}^{(t)},\mat{p}_{\mat{X}}^{(t)}}\mat{p}_{\mat{A}}^{(t)'}\mat{Y}_{\mat{A}}^{(t)'}\mat{Y}_{\mat{A}}^{(t)}\mat{p}_{\mat{A}}^{(t)}+\mat{p}_{\mat{A}}^{(t)'}\mat{Y}_{\mat{A}}^{(t)'}\mat{Y}_{\mat{X}}^{(t)}\mat{p}_{\mat{X}}^{(t)}\\
&+\mat{p}_{\mat{X}}^{(t)'}\mat{Y}_{\mat{X}}^{(t)'}\mat{Y}_{\mat{A}}^{(t)}\mat{p}_{\mat{A}}^{(t)}+\mat{p}_{\mat{X}}^{(t)'}\mat{Y}_{\mat{X}}^{(t)'}\mat{Y}_{\mat{X}}^{(t)}\mat{p}_{\mat{X}}^{(t)}.\\
&\mbox{s.t.}\quad  \mat{p}_{\mat{A}}^{(t)'}\mat{Y}_{\mat{A}}^{(t)'}\mat{Y}_{\mat{A}}^{(t)}\mat{p}_{\mat{A}}^{(t)}+\mat{p}_{\mat{X}}^{(t)'}\mat{Y}_{\mat{X}}^{(t)'}\mat{Y}_{\mat{X}}^{(t)}\mat{p}_{\mat{X}}^{(t)}=1.
\end{split}
\end{equation}

Let $\gamma$ be the Lagrange multiplier for the constraint, by setting the derivative of the Lagrange function w.r.t. $\mat{p}_{\mat{A}}^{(t)}$ and $\mat{p}_{\mat{X}}^{(t)}$ to zero, we obtain the optimal solution for $[\mat{p}_{\mat{A}}^{(t)};\mat{p}_{\mat{X}}^{(t)}]$, which corresponds to the eigenvector of the following generalized eigen-problem:
\begin{equation}
\begin{split}
\begin{bmatrix}
    \mat{Y}_{\mat{A}}^{(t)'}\mat{Y}_{\mat{A}}^{(t)} & \mat{Y}_{\mat{A}}^{(t)'}\mat{Y}_{\mat{X}}^{(t)} \\
    \mat{Y}_{\mat{X}}^{(t)'}\mat{Y}_{\mat{A}}^{(t)} & \mat{Y}_{\mat{X}}^{(t)'}\mat{Y}_{\mat{X}}^{(t)} \\
\end{bmatrix}
\begin{bmatrix}
    \mat{p}_{\mat{A}}^{(t)} \\
    \mat{p}_{\mat{X}}^{(t)} \\
\end{bmatrix}
=
\gamma&
\begin{bmatrix}
    \mat{Y}_{\mat{A}}^{(t)'}\mat{Y}_{\mat{A}}^{(t)} & \mat{0} \\
    \mat{0} & \mat{Y}_{\mat{X}}^{(t)'}\mat{Y}_{\mat{X}}^{(t)} \\
\end{bmatrix}
\begin{bmatrix}
    \mat{p}_{\mat{A}}^{(t)} \\
    \mat{p}_{\mat{X}}^{(t)} \\
\end{bmatrix}.
\end{split}
\label{eq:embeddingfushion}
\end{equation}As a result, to obtain a consensus embedding representation from $\mat{Y}_{\mat{A}}$ and $\mat{Y}_{\mat{X}}$, we could take the top-$l$ eigenvectors of the above generalized eigen-problem and stack these top-$l$ eigenvectors together. Suppose the projection matrix $\mat{P}^{(t)}\in\mathbb{R}^{2k\times l}$ is the concatenated top-$l$ eigenvectors, the final consensus embedding representation can be computed as $\mat{Y}^{(t)}=[\mat{Y}_{\mat{A}}^{(t)},\mat{Y}_{\mat{X}}^{(t)}]\times\mat{P}^{(t)}$.

\subsection{Online Model of DANE}
More often than not, attributed networks often exhibit high dynamics. For example, in social media sites, social relations are continuously evolving, and user posting behaviors may also evolve accordingly. It raises challenges to the existing offline embedding methods as they have to rerun at each time step which is time-consuming and is not scalable to large networks. Therefore, it is important to build an efficient online embedding algorithm which gives an informative embedding representation on the fly.

The proposed online embedding model is motivated by the observation that most of real-world networks, with no exception for attributed networks, often evolve smoothly in the temporal dimension between two consecutive time steps~\cite{chi2007evolutionary,aggarwal2014evolutionary,wang2016recommending,li2016toward}. Hence, we use $\Delta\mat{A}$ and $\Delta\mat{X}$ to denote the perturbation of network structure and node attributes between two consecutive time steps $t$ and $t+1$, respectively. With these, the diagonal matrix and Laplacian matrix of $\mat{A}$ and $\mat{X}$ also evolve smoothly such that:
\begin{equation}
\begin{split}
\mat{D}_{\mat{A}}^{(t+1)} &= \mat{D}_{\mat{A}}^{(t)}+\Delta\mat{D}_{\mat{A}}, \quad \mat{L}_{\mat{A}}^{(t+1)} = \mat{L}_{\mat{A}}^{(t)}+\Delta\mat{L}_{\mat{A}},\\
\mat{D}_{\mat{X}}^{(t+1)} &= \mat{D}_{\mat{X}}^{(t)}+\Delta\mat{D}_{\mat{X}}, \quad \mat{L}_{\mat{X}}^{(t+1)} = \mat{L}_{\mat{X}}^{(t)}+\Delta\mat{L}_{\mat{X}}.
\end{split}
\end{equation}

As discussed in the previous subsection, the problem of attributed network embedding in an offline setting boils down to solving generalized eigen-problems. In particular, offline model focuses on finding the top eigenvectors corresponding to the smallest eigenvalues of the generalized eigen-problems. Therefore, the core idea to enable online update of the embeddings is to develop an efficient way to update the top eigenvectors and eigenvalues. Otherwise, we have to perform generalized eigen-decomposition each time step, which is not practical due to its high time complexity.

Without loss of generality, we use the network topology as an example to illustrate the proposed algorithm for online embedding. By the matrix perturbation theory~\cite{stewart1990matrix}, we have the following equation in embedding the network structure at the new time step:
\begin{equation}
(\mat{L}_{\mat{A}}^{(t)}+\Delta\mat{L}_{\mat{A}})(\mat{a}+\Delta\mat{a})=(\lambda+\Delta\lambda)(\mat{D}_{\mat{A}}^{(t)}+\Delta\mat{D}_{\mat{A}})(\mat{a}+\Delta\mat{a}).
\end{equation}For a specific eigen-pair $(\lambda_{i},\mat{a}_{i})$, we have the following equation:
\begin{equation}
(\mat{L}_{\mat{A}}^{(t)}+\Delta\mat{L}_{\mat{A}})(\mat{a}_{i}+\Delta\mat{a}_{i})=(\lambda_{i}+\Delta\lambda_{i})(\mat{D}_{\mat{A}}^{(t)}+\Delta\mat{D}_{\mat{A}})(\mat{a}_{i}+\Delta\mat{a}_{i}).
\end{equation}The problem now is how to compute the change of the $i$-th eigen-pair $(\Delta\mat{a}_{i}, \Delta\lambda_{i})$ by taking advantage of the small perturbation matrices $\Delta\mat{D}$ and $\Delta\mat{L}$.
\paragraph{\textbf{A - Computing the change of eigenvalue $\mat{\Delta}\mat{\lambda_{i}}$}}~\\
By expanding the above equation, we have:
\begin{equation}
\begin{split}
&\mat{L}_{\mat{A}}^{(t)}\mat{a}_{i}+\Delta\mat{L}_{\mat{A}}\mat{a}_{i}+\mat{L}_{\mat{A}}^{(t)}\Delta\mat{a}_{i}+\Delta\mat{L}_{\mat{A}}\Delta\mat{a}_{i}\\
=&\lambda_{i}\mat{D}_{\mat{A}}^{(t)}\mat{a}_{i}+\lambda_{i}\Delta\mat{D}_{\mat{A}}\mat{a}_{i}+\Delta\lambda_{i}\mat{D}_{\mat{A}}^{(t)}\mat{a}_{i}+\Delta\lambda_{i}\Delta\mat{D}_{\mat{A}}\mat{a}_{i}\\
+&(\lambda_{i}\mat{D}_{\mat{A}}^{(t)}+\lambda_{i}\Delta\mat{D}_{\mat{A}}+\Delta\lambda_{i}\mat{D}_{\mat{A}}^{(t)}+\Delta\lambda_{i}\Delta\mat{D}_{\mat{A}})\Delta\mat{a}_{i}.
\end{split}
\label{eq:expasion}
\end{equation}The higher order terms, i.e., $\Delta\lambda_{i}\Delta\mat{D}_{\mat{A}}\mat{a}_{i}$, $\lambda_{i}\Delta\mat{D}_{\mat{A}}\Delta\mat{a}_{i}$, $\Delta\lambda_{i}\mat{D}_{\mat{A}}^{(t)}\Delta\mat{a}_{i}$ and $\Delta\lambda_{i}\Delta\mat{D}_{\mat{A}}\Delta\mat{a}_{i}$ can be removed as they have limited effects on the accuracy of the generalized eigen-systems~\cite{golub2012matrix}. By using the fact that $\mat{L}_{\mat{A}}^{(t)}\mat{a}_{i}=\lambda_{i}\mat{D}_{\mat{A}}^{(t)}\mat{a}_{i}$, we have the following formulation:
\begin{equation}
\Delta\mat{L}_{\mat{A}}\mat{a}_{i}+\mat{L}_{\mat{A}}^{(t)}\Delta\mat{a}_{i}=\lambda_{i}\Delta\mat{D}_{\mat{A}}\mat{a}_{i}+\Delta\lambda_{i}\mat{D}_{\mat{A}}^{(t)}\mat{a}_{i}+\lambda_{i}\mat{D}_{\mat{A}}^{(t)}\Delta\mat{a}_{i}.
\label{eq:firstorder}
\end{equation}Multiplying both sides with $\mat{a}_{i}'$, we now have:
\begin{equation}
\mat{a}_{i}'\Delta\mat{L}_{\mat{A}}\mat{a}_{i}+\mat{a}_{i}'\mat{L}_{\mat{A}}^{(t)}\Delta\mat{a}_{i}=\lambda_{i}\mat{a}_{i}'\Delta\mat{D}_{\mat{A}}\mat{a}_{i}+\Delta\lambda_{i}\mat{a}_{i}'\mat{D}_{\mat{A}}^{(t)}\mat{a}_{i}+
\lambda_{i}\mat{a}_{i}'\mat{D}_{\mat{A}}^{(t)}\Delta\mat{a}_{i}.
\label{eq:transpose}
\end{equation}Since both the Laplacian matrix $\mat{L}_{\mat{A}}^{(t)}$ and the diagonal matrix $\mat{D}_{\mat{A}}^{(t)}$ are symmetric, we have:
\begin{equation}
\mat{a}_{i}'\mat{L}_{\mat{A}}^{(t)}\Delta\mat{a}_{i}=\lambda_{i}\mat{a}_{i}'\mat{D}_{\mat{A}}^{(t)}\Delta\mat{a}_{i}.
\end{equation}Therefore, Eq.~(\ref{eq:transpose}) can be reformulated as follows:
\begin{equation}
\mat{a}_{i}'\Delta\mat{L}_{\mat{A}}\mat{a}_{i}=\lambda_{i}\mat{a}_{i}'\Delta\mat{D}_{\mat{A}}\mat{a}_{i}+\Delta\lambda_{i}\mat{a}_{i}'\mat{D}_{\mat{A}}^{(t)}\mat{a}_{i}.
\end{equation}Through this, the variation of eigenvalue, i.e., $\Delta\lambda_{i}$, is:
\begin{equation}
\Delta\lambda_{i}=\frac{\mat{a}_{i}'\Delta\mat{L}_{\mat{A}}\mat{a}_{i}-\lambda_{i}\mat{a}_{i}'\Delta\mat{D}_{\mat{A}}\mat{a}_{i}}{\mat{a}_{i}'\mat{D}_{\mat{A}}^{(t)}\mat{a}_{i}}.
\label{eq:eigenvaluesolution}
\end{equation}
\begin{theorem} In the generalized eigen-problem $\mat{A}\mat{v}=\lambda\mat{B}\mat{v}$, if $\mat{A}$ and $\mat{B}$ are both Hermitian matrices and $\mat{B}$ is a positive-semidefinite matrix, the eigenvalue $\lambda$ are real; and eigenvectors $\mat{v}_{j}$ ($i\neq j$) are $\mat{B}$-orthogonal such that $\mat{v}_{i}'\mat{B}\mat{v}_{j}=0$ and $\mat{v}_{i}'\mat{B}\mat{v}_{i}=1$~\cite{parlett1980symmetric}.
\label{theorem:theorem1}
\end{theorem}

\begin{corollary}
$\mat{a}_{i}'\mat{D}_{\mat{A}}^{(t)}\mat{a}_{i}=1$ and $\mat{a}_{i}'\mat{D}_{\mat{A}}^{(t)}\mat{a}_{j}=0$ ($i\neq j$).
\label{corollary:corollary1}
\end{corollary}
\label{corollary:orthonormal}
\begin{proof}
Both $\mat{D}_{\mat{A}}^{(t)}$ and $\mat{L}_{\mat{A}}^{(t)}$ are symmetric and are also Hermitian matrices. Meanwhile, the Laplacian matrix $\mat{L}_{\mat{A}}^{(t)}$ is a positive-definite matrix, which completes the proof.
\end{proof}Therefore, the variation of the eigenvalue $\lambda_{i}$ is as follows:
\begin{equation}
\Delta\lambda_{i}=\mat{a}_{i}'\Delta\mat{L}_{\mat{A}}\mat{a}_{i}-\lambda_{i}\mat{a}_{i}'\Delta\mat{D}_{\mat{A}}\mat{a}_{i}.
\label{eq:eigenvaluefinalupdate}
\end{equation}

\paragraph{\textbf{B - Computing the change of eigenvector $\mat{\Delta}\mat{a}_{i}$}}~\\
As network structure often evolves smoothly between two continuous time steps, we assume that the perturbation of the eigenvectors $\Delta\mat{a}_{i}$ lies in the column space that is composed by the top-$k$ eigenvectors at time step $t$ such that $\Delta\mat{a}_{i}=\sum_{j=2}^{k+1}\alpha_{ij}\mat{a}_{j}$, where $\alpha_{ij}$ is a weight indicating the contribution of the $j$-th eigenvector $\mat{a}_{j}$ in approximating the new $i$-th eigenvector. Next, we show how to determine these weights such that the perturbation $\Delta\mat{a}_{i}$ can be estimated.

By plugging $\Delta\mat{a}_{i}=\sum_{j=2}^{k+1}\alpha_{ij}\mat{a}_{j}$ into Eq.~(\ref{eq:firstorder}) and using the fact that $\mat{L}_{\mat{A}}^{(t)}\sum_{j=2}^{k+1}\alpha_{ij}\mat{a}_{j}=\mat{D}_{\mat{A}}^{(t)}\sum_{j=2}^{k+1}\alpha_{ij}\lambda_{j}\mat{a}_{j}$, we obtain the following:
\begin{equation}
\small
\Delta\mat{L}_{\mat{A}}\mat{a}_{i}+\mat{D}_{\mat{A}}^{(t)}\sum_{j=2}^{k+1}\alpha_{ij}\lambda_{j}\mat{a}_{j}=\lambda_{i}\Delta\mat{D}_{\mat{A}}\mat{a}_{i}+\Delta\lambda_{i}\mat{D}_{\mat{A}}^{(t)}\mat{a}_{i}+\lambda_{i}\mat{D}_{\mat{A}}^{(t)}\sum_{j=2}^{k+1}\alpha_{ij}\mat{a}_{j}.
\label{eq:eigenvectornew2}
\end{equation}By multiplying eigenvector $\mat{a}_{p}'\,(2\leq p\leq k+1, p\neq i)$ on both sides of Eq.~(\ref{eq:eigenvectornew2}) and taking advantage of the orthonormal property from Corollary~\ref{corollary:orthonormal}, we obtain the following:
\begin{equation}
\begin{split}
&\mat{a}_{p}'\Delta\mat{L}_{\mat{A}}\mat{a}_{i}+\mat{a}_{p}'\mat{D}_{\mat{A}}^{(t)}\sum_{j=2}^{k+1}\alpha_{ij}\lambda_{j}\mat{a}_{j}\\
=\,&\lambda_{i}\mat{a}_{p}'\Delta\mat{D}_{\mat{A}}\mat{a}_{i}+\Delta\lambda_{i}\mat{a}_{p}'\mat{D}_{\mat{A}}^{(t)}\mat{a}_{i}+\lambda_{i}\mat{a}_{p}'\mat{D}_{\mat{A}}^{(t)}\sum_{j=2}^{k+1}\alpha_{ij}\mat{a}_{j}\\
\Rightarrow \quad &\mat{a}_{p}'\Delta\mat{L}_{\mat{A}}\mat{a}_{i}+\alpha_{ip}\lambda_{p}=\lambda_{i}\mat{a}_{p}'\Delta\mat{D}_{\mat{A}}\mat{a}_{i}+\alpha_{ip}\lambda_{i}.
\end{split}
\end{equation}Hence, the weight $\alpha_{ip}$ can be determined by:
\begin{equation}
\alpha_{ip}=\frac{\mat{a}_{p}'\Delta\mat{L}_{\mat{A}}\mat{a}_{i}-\lambda_{i}\mat{a}_{p}'\Delta\mat{D}_{\mat{A}}\mat{a}_{i}}{\lambda_{i}-\lambda_{p}}.
\label{eq:alphaip}
\end{equation}After eigenvector perturbation, we still need to make the orthonormal condition holds for new eigenvectors, thus we have $(\mat{a}_{i}+\Delta\mat{a}_{i})'(\mat{D}_{\mat{A}}+\Delta\mat{D}_{\mat{A}})(\mat{a}_{i}+\Delta\mat{a}_{i})=1$. By expanding it and removing the second-order and third-order terms, we obtain the following equation:
\begin{equation}
2\mat{a}_{i}'\mat{D}_{\mat{A}}^{(t)}\Delta\mat{a}_{i}+\mat{a}_{i}'\Delta\mat{D}_{\mat{A}}^{(t)}\mat{a}_{i}=0.
\label{eq:alphaiitmp}
\end{equation}Then the solution of $\alpha_{ii}$ is as follows:
\begin{equation}
\alpha_{ii}=-\frac{1}{2}\mat{a}_{i}'\Delta\mat{D}_{\mat{A}}\mat{a}_{i}.
\label{eq:alphaii}
\end{equation}

With the solutions of $\alpha_{ip}$ ($p\neq i$) and $\alpha_{ii}$, the perturbation of eigenvector $\mat{a}_{i}$ is given as follows:
\begin{equation}
\Delta\mat{a}_{i}=-\frac{1}{2}\mat{a}_{i}'\Delta\mat{D}_{\mat{A}}\mat{a}_{i}\mat{a}_{i}+\sum_{j=2,j\neq i}^{k+1}(\frac{\mat{a}_{j}'\Delta\mat{L}_{\mat{A}}\mat{a}_{i}-\lambda_{i}\mat{a}_{j}'\Delta\mat{D}_{\mat{A}}\mat{a}_{i}}{\lambda_{i}-\lambda_{j}})\mat{a}_{j}.
\label{eq:eigenvectorfinalupdate}
\end{equation}

Overall, the $i$-th eigen-pair ($\Delta\lambda_{i}, \Delta\mat{a}_{i}$) can be updated on the fly by Eq.~(\ref{eq:eigenvaluefinalupdate}) and Eq.~(\ref{eq:eigenvectorfinalupdate}), the pseudocode of the updating process is illustrated in Algorithm~\ref{alg:generalizedeigenupdate}. The first input is the top-$k$ eigen-pairs of the generalized eigen-problem, they can be computed by standard methods like power iteration and Lanczos method~\cite{golub2012matrix}. Another input is the variation of the diagonal matrix and the Laplacian matrix. For the top-$k$ eigen-pairs, we update eigenvalues in line 2 and update eigenvectors in line 3.

Likewise, the embedding of node attributes can also be updated in an online manner by Algorithm~\ref{alg:generalizedeigenupdate}. Specifically, let $\mat{Y}_{\mat{A}}^{(t)}$ and $\mat{Y}_{\mat{X}}^{(t)}$ denote the embedding of network structure and node attributes at time step $t$, then at the following time step $t+1$, we first employ the proposed online model to update their embedding representations, then a final consensus embedding representation $\mat{Y}^{(t+1)}$ is derived by the correlation maximization method mentioned previously.

\begin{algorithm}[!htbp]
\begin{algorithmic}[1]
    \Require Top-$k$ eigen-pairs of the generalized eigen-problem $\{$($\lambda_{2},\mat{a}_{2}$),($\lambda_{3},\mat{a}_{3}$),...,($\lambda_{k+1},\mat{a}_{k+1}$)$\}$ at time $t$, variation of the diagonal matrix $\Delta\mat{L}_{\mat{A}}$ and Laplacian matrix $\Delta\mat{D}_{\mat{A}}$.
    \Ensure Top-$k$ eigen-pairs $\{$($\lambda_{2}^{(t+1)},\mat{a}_{2}^{(t+1)}$),...,($\lambda_{k+1}^{(t+1)},\mat{a}_{k+1}^{(t+1)}$)$\}$ at time step $t+1$.
    \For {$i=2$ to $k+1$}
        \State {Calculate the variation of $\Delta\lambda_{i}$ by Eq.~(\ref{eq:eigenvaluefinalupdate});}
        \State {Calculate the variation of $\Delta\mat{a}_{i}$ by Eq.~(\ref{eq:eigenvectorfinalupdate});}
        \State {$\lambda_{i}^{(t+1)}=\lambda_{i}+\Delta\lambda_{i}$; $\mat{a}_{i}^{(t+1)}=\mat{a}_{i}+\Delta\mat{a}_{i}$;}
    \EndFor
\end{algorithmic}
\caption{Updating of embedding results for the network}
\label{alg:generalizedeigenupdate}
\end{algorithm}

\paragraph{\textbf{C - Computational Complexity Analysis}}
We theoretically analyze the computational complexity of the proposed online algorithm and show its superiority over the offline embedding methods.
\begin{lemma}
The time complexity of the proposed online embedding algorithm over $T$ time steps is $\mathcal{O}(Tk^{2}(n+l+l_{a}+l_{x}+d_{x}+d_{x}))$, where $k$ is the intermediate embedding dimension for network (or attributes), $l$ is the final consensus embedding dimension, $n$ is the number of nodes, and $l_{a}$, $l_{x}$, $d_{a}$, $d_{x}$ are the number of non-zero entries in the sparse matrices $\mat{\Delta}\mat{L}_{\mat{A}}$, $\mat{\Delta}\mat{L}_{\mat{X}}$, $\mat{\Delta}\mat{D}_{\mat{A}}$, and $\mat{\Delta}\mat{D}_{\mat{X}}$, respectively.
\begin{proof}
In each time step, to update the top-$k$ eigenvalues of the network and node attributes in an online fashion, it requires $\mathcal{O}(k(d_{a}+l_{a}))$ and $\mathcal{O}(k(d_{x}+l_{x}))$, respectively. Also, the online updating of the top-$k$ eigenvectors for the network and attributes are $\mathcal{O}(k^{2}(d_{a}+l_{a}+n))$ and $\mathcal{O}(k^{2}(d_{x}+l_{x}+n))$, respectively. After that, the complexity for the consensus embedding is $\mathcal{O}(k^{2}l)$. Therefore, the computational complexity of the proposed online model over $T$ time steps are $\mathcal{O}(Tk^{2}(n+l+l_{a}+l_{x}+d_{x}+d_{x}))$.
\end{proof}
\end{lemma}

\begin{lemma}
The time complexity of the proposed offline embedding algorithm over $T$ time steps is $\mathcal{O}(Tn^{2}(k+l))$, where $k$ is the intermediate embedding dimension for network (or attributes), $l$ is the final consensus embedding dimension.
\begin{proof}
Omitted for brevity.
\end{proof}
\end{lemma}

As can be shown, since $\mat{\Delta}\mat{L}_{\mat{A}}$, $\mat{\Delta}\mat{L}_{\mat{X}}$, $\mat{\Delta}\mat{D}_{\mat{A}}$, and $\mat{\Delta}\mat{D}_{\mat{X}}$ are often very sparse, thus $l_{a}$, $l_{x}$, $d_{a}$, $d_{x}$ are usually very small, meanwhile we have $k\ll n$ and $l \ll n$. Based on the above analysis, the proposed online embedding algorithm for dynamic attributed networks is much more efficient than rerunning the offline method repeatedly.

\section{Experiments}
In this section, we conduct experiments to evaluate the effectiveness and efficiency of the proposed DANE framework for dynamic attributed network embedding. In particular, we attempt to answer the following two questions: (1) \emph{Effectiveness}:  how effective are the embeddings obtained by DANE on different learning tasks? (2) \emph{Efficiency}: how fast is the proposed framework DANE compared with other offline embedding methods? We first introduce the datasets and experimental settings before presenting details of the experimental results.
\subsection{Datasets}
We use four datasets BlogCatalog, Flickr, Epinions and DBLP for experimental evaluation. Among them, BlogCatalog and Flickr are synthetic data from static attributed networks, and they have been used in previous research~\cite{li2015unsupervised,li2016robust}. We randomly add 0.1\% new edges and change 0.1\% attribute values at each time step to simulate its evolving nature. The other two datasets, Epinions and DBLP are real-world dynamic attributed networks. Epinions is a product review site in which users share their reviews and opinions about products. Users themselves can also build trust networks to seek advice from others. Node attributes are formed by the bag-of-words model on the reviews, while the major categories of reviews by users are taken as the ground truth of class labels. The data has 16 different time steps. In the last dataset DBLP, we extracted a DBLP co-author network for the authors that publish at least two papers between the years of 2001 and 2016 from seven different areas. Bag-of-words model is applied on the paper title to obtain the attribute information, and the major area the authors publish is considered as ground truth. It should be noted that in all these four datasets, the evolution of network structure and node attributes are very smooth. The detailed statistics of these datasets are listed in Table~\ref{table:datasets}.
\begin{table}
\centering
\begin{tabular}{c|c|c|c|c} \hline
& BlogCatalog & Flickr & Epinions &  DBLP\\ \hline \hline
$\#$ Nodes & 5,196  & 7,575 & 14,180 & 23,393\\ \hline
$\#$ Attributes  & 8,189  & 12,047 & 9,936 & 8,945 \\ \hline
$\#$ Edges & 173,468 & 242,146 & 227,642 & 289,478 \\ \hline
$\#$ Classes &  6 & 9 & 20 & 7\\ \hline
$\#$ Time Steps & 10 & 10 & 16 & 16 \\ \hline
\end{tabular}
\caption{Detailed information of the datasets.}
\label{table:datasets}
\end{table}

\subsection{Experimental Settings}
One commonly adopted way to evaluate the quality of the embedding representation~\cite{chang2015heterogeneous,jacob2014learning,perozzi2014deepwalk,tang2015line} is by the following two unsupervised and supervised tasks: network clustering and node classification. First, we validate the effectiveness of the embedding representations by DANE on the network clustering task. Two standard clustering performance metrics, i.e., \emph{clustering accuracy} (ACC) and \emph{normalized mutual information} (NMI) are used. In particular, after obtaining the embedding representation of each node in the attributed network, we perform K-means clustering based on the embedding representations. The K-means algorithm is repeated 10 times and the average results are reported since K-means may converge to the local minima due to different initializations. Another way to assess the embedding is by the node classification task. Specifically, we split the the embedding representations of all nodes via a 10-fold cross-validation, using 90\% of nodes to train a classification model by logistic regression and the rest 10\% nodes for the testing. The whole process is repeated 10 times and the average performance are reported. Three evaluation metrics, \emph{classification accuracy}, \emph{F1-Macro} and \emph{F1-Micro} are used. How to determine the optimal number of embedding dimensions is still an open research problem, thus we vary the embedding dimension as $\{10,20,...,100\}$ and the best results are reported.

\subsubsection{Baseline Methods}
DANE is measured against the following baseline methods on the two aforementioned tasks:
\begin{itemize}
\item \textbf{Deepwalk}: learns network embeddings by word2vec and truncated random walk techniques~\cite{perozzi2014deepwalk}.
\item \textbf{LINE}: learns embeddings by preserving the first-order and second-order proximity structures of the network~\cite{tang2015line}.
\item \textbf{DANE-N}: is a variation of the proposed DANE with only network information.
\item \textbf{DANE-A}: is a variation of the proposed DANE with only attribute information.
\item \textbf{CCA}: directly uses the original network structure and attributes for a joint low-dimensional representation~\cite{hardoon2004canonical}.
\item \textbf{LCMF}: maps network and attributes to a shared latent space by collective matrix factorization~\cite{zhu2007combining}.
\item \textbf{LANE}: is a label informed attributed network embedding method, we use one of its variant LANE w/o Label~\cite{huang2017label}.
\item \textbf{DANE-O}: is a variation of DANE that reruns the offline model at each time step.
\end{itemize}

It is important to note that Deepwalk, LINE, CCA, LCMF, LANE, and DANE-O can only handle static networks. To have a fair comparison with the proposed DANE framework, we rerun these baseline methods at each time step and report the average performance over all time steps\footnote{For baseline methods that cannot finish in 24hrs, we only run it once. As networks evolve smoothly, there is not much difference in terms of average performance.}. We follow the suggestions of the original papers to set the parameters of all these baselines.

\subsection{Unsupervised Task - Network Clustering}
To evaluate the effectiveness of embedding representations, we first compare DANE with baseline methods on network clustering which is naturally an unsupervised learning task. As per the fact that the attributed networks are constantly evolving, we compare the average clustering performance over all time steps. The average clustering performance comparison w.r.t. ACC and NMI are presented in Table~\ref{table:clustering}. We make the following observations:

\begin{table*}[!t]
\centering
\caption{Clustering results ($\%$) comparison of different embedding methods.}
\newcommand{\minitab}[2][l]{\begin{tabular}{#1}#2\end{tabular}}
\begin{tabular}{|c|c||c|c||c|c||c|c||c|c|}\hline
\multicolumn{2}{|c||}{Datasets}  & \multicolumn{2}{|c||}{BlogCatalog}  &   \multicolumn{2}{|c||}{Flickr}   & \multicolumn{2}{|c||}{Epinions} & \multicolumn{2}{|c|}{DBLP}      \\ \hline\hline
\multicolumn{2}{|c||}{Methods}  & ACC & NMI & ACC & NMI & ACC & NMI & ACC & NMI  \\ \hline
\multirow{3}{*}{Network} & Deepwalk & 49.85 & 30.51 & 40.70 & 24.29 & 13.31 & 12.72 & 53.61 & 32.54 \\ \cline{2-10}
& LINE     & 50.20 & 29.53 & 42.93 & 26.01 & 14.34 & 12.65 & 51.61 & 30.74 \\ \cline{2-10}
 & DANE-N	 & 37.05 & 21.84 & 31.89 & 18.91 & 12.01 & 11.95 & 56.61 & 31.54 \\ \hline \hline
Attributes & DANE-A	 & 62.32 & 45.95 & 63.80 & 48.29 & 16.12 & 11.62 & 47.37 & 20.64 \\ \hline \hline
\multirow{5}{*}{Network+Attributes} & CCA	     & 33.42 & 11.86 & 24.39 & 10.89 & 10.85 &  8.61 & 26.42 & 18.60 \\ \cline{2-10}
&LCMF	 & 55.72 & 40.38 & 27.03 & 13.06 & 12.86 & 10.73 & 42.27 & 26.48 \\ \cline{2-10}
&LANE	 & 65.06 & 48.89 & 65.45 & 52.58 & 32.18 & 22.09 & 55.80 & 31.84 \\ \cline{2-10}
&DANE-O	 & 80.31 & 59.46 & 67.33 & 53.04 & 34.11 & 23.07 & 59.14 & 35.31 \\ \cline{2-10}
&DANE	 & 79.69 & 59.32 & 67.24 & 52.19 & 34.52 & 22.36 & 57.68 & 34.87 \\ \hline
\end{tabular}
\label{table:clustering}
\end{table*}

\begin{itemize}
\item DANE and its offline version DANE-O consistently outperform all baseline methods on four dynamic attributed networks by achieving better clustering performance. We also perform pairwise Wilcoxon signed-rank test~\cite{demvsar2006statistical} between DANE, DANE-O and these baseline methods and the test results show that DANE and DANE-O are significantly better (with both 0.01 and 0.05 significance levels).
\item DANE, DANE-O and LANE achieve better clustering performance than network embedding methods such as Deepwalk, LINE and DANE-N and attribute embedding method DANE-A. The improvements indicate that attribute information is complementary to pure network topology and can help learn more informative embedding representations. Meanwhile, DANE also outperforms the CCA and LCMF which also leverage node attributes. The reason is that although these methods learn a low-dimensional representation by using both sources, they are not explicitly designed to preserve the node proximity. Also, their performance degenerates when the data is very noisy.
\item Even though DANE leverages matrix perturbation theory to update the embedding representations, its performance is very close to DANE-O which reruns at each time step. It implies that the online embedding model does not sacrifice too much informative information in terms of embedding.
\end{itemize}

\subsection{Supervised Task - Node Classification}
\begin{table*}[!t]
\centering
\caption{Classification results ($\%$) comparison of different embedding methods.}
\newcommand{\minitab}[2][l]{\begin{tabular}{#1}#2\end{tabular}}
\begin{tabular}{|c|c||c|c|c||c|c|c||c|c|c||c|c|c|}\hline
\multicolumn{2}{|c||}{Datasets}        & \multicolumn{3}{|c||}{BlogCatalog}  &   \multicolumn{3}{|c||}{Flickr}   & \multicolumn{3}{|c||}{Epinions} & \multicolumn{3}{|c|}{DBLP}      \\ \hline\hline
\multicolumn{2}{|c||}{Methods}  & AC & Micro & Macro & AC & Micro & Macro & AC & Micro & Macro & AC & Micro & Macro \\ \hline
\multirow{3}{*}{Network} & Deepwalk & 68.05 & 67.15 & 68.18 & 60.08 & 58.93 & 59.08 & 22.12 & 17.43 & 20.10 & 74.38 & 69.65 & 72.37 \\ \cline{2-14}
&LINE     & 70.20 & 69.88 & 70.91 & 61.03 & 60.90 & 60.01 & 23.54 & 17.17 & 21.05 & 72.97 & 67.56 & 70.97 \\ \cline{2-14}
&DANE-N   & 66.97 & 66.06 & 67.78 & 49.37 & 47.82 & 49.34 & 21.25 & 20.57 & 21.88 & 71.99 & 65.33 & 71.94 \\ \hline \hline
Attributes&DANE-A   & 80.23 & 79.86 & 80.23 & 76.66 & 75.59 & 76.60 & 23.76 & 21.57 & 22.00 & 63.92 & 54.80 & 62.97 \\ \hline \hline
\multirow{5}{*}{Network+Attributes} &CCA      & 48.63 & 49.96 & 49.63 & 27.09 & 26.54 & 26.09 & 11.53 &  9.43 & 10.56 & 45.67 & 42.08 & 43.83 \\ \cline{2-14}
&LCMF	 & 84.41 & 89.01 & 89.26 & 66.27 & 66.75 & 65.71 & 19.14 &  9.22 & 10.14 & 69.71 & 68.01 & 68.42 \\ \cline{2-14}
&LANE     & 87.52 & 87.52 & 87.93 & 77.54 & 77.81 & 77.26 & 27.74 & 28.45 & 28.87 & 72.15 & 71.09 & 73.48 \\ \cline{2-14}
&DANE-O   & 89.34 & 89.15 & 89.23 & 79.68 & 79.52 & 79.95 & 31.23 & 31.28 & 31.35 & 77.21 & 74.96 & 75.48 \\ \cline{2-14}
&DANE     & 89.09 & 88.78 & 88.94 & 79.56 & 78.94 & 79.56 & 30.87 & 30.93 & 30.81 & 76.64 & 74.53 & 75.69 \\ \hline
\end{tabular}
\label{table:classification}
\end{table*}
Next, we assess the effectiveness of embedding representations on a supervised learning task - node classification. Similar to the settings of network clustering, we report the average classification performance over all time steps. The classification results in terms of three different measures are shown in Table~\ref{table:classification}. The following findings can be inferred from the table:
\begin{itemize}
\item Generally, we have the similar observations as the clustering task. The methods which only use link information or node attributes (e.g., Deepwalk, LINE, DANE-N, DANE-A) and methods which do not explicitly model node proximity (e.g., CCA, LCMF) give poor classification results.
\item The embeddings learned by DANE and DANE-O help train a more discriminative classification model by obtaining higher classification performance. In addition, pairwise Wilcoxon signed-rank test~\cite{demvsar2006statistical} shows that DANE and DANE-O are significantly better.
\item For the node classification task, the attribute embedding method DANE-A works better than the network embedding method in the BlogCatalog, Flickr and Epinions datasets. The reason is that in these datasets, the class labels are more closely related to the attribute information than the network structure. However, it is a different case for the DBLP dataset in which the labels of authors are more closely related to the coauthor relationships.
\end{itemize}

\subsection{Efficiency of Online Embedding}
\begin{figure*}[!t]
\centering
\begin{minipage}{0.48\textwidth}
\centering
\subfigure[BlogCatalog\label{fig:blogcatalogcumulative}]
{\includegraphics[width=\textwidth]{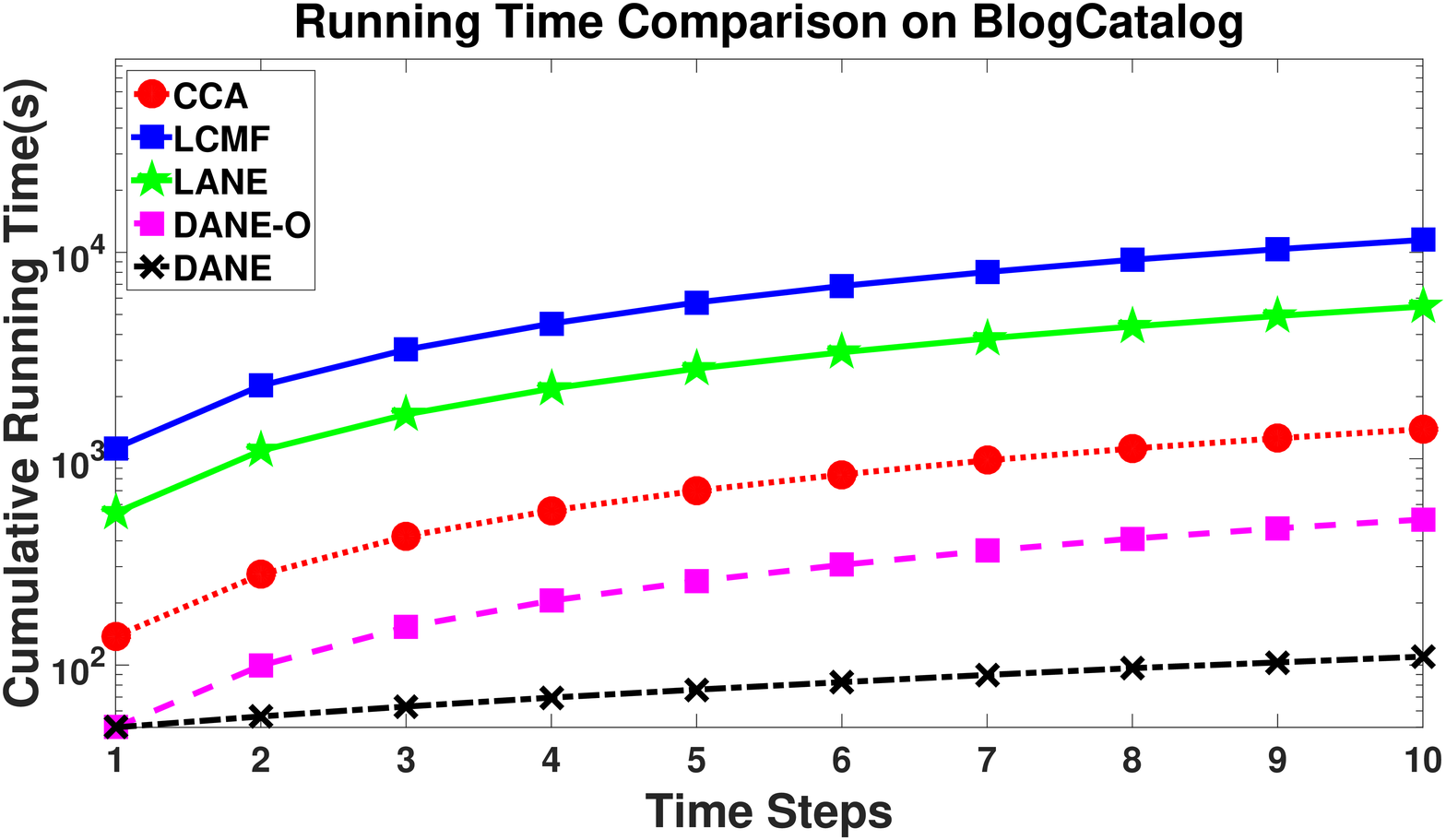}}
\end{minipage}
\begin{minipage}{0.48\textwidth}
\centering
\subfigure[Flickr\label{fig:flickrcumulative}]
{\includegraphics[width=\textwidth]{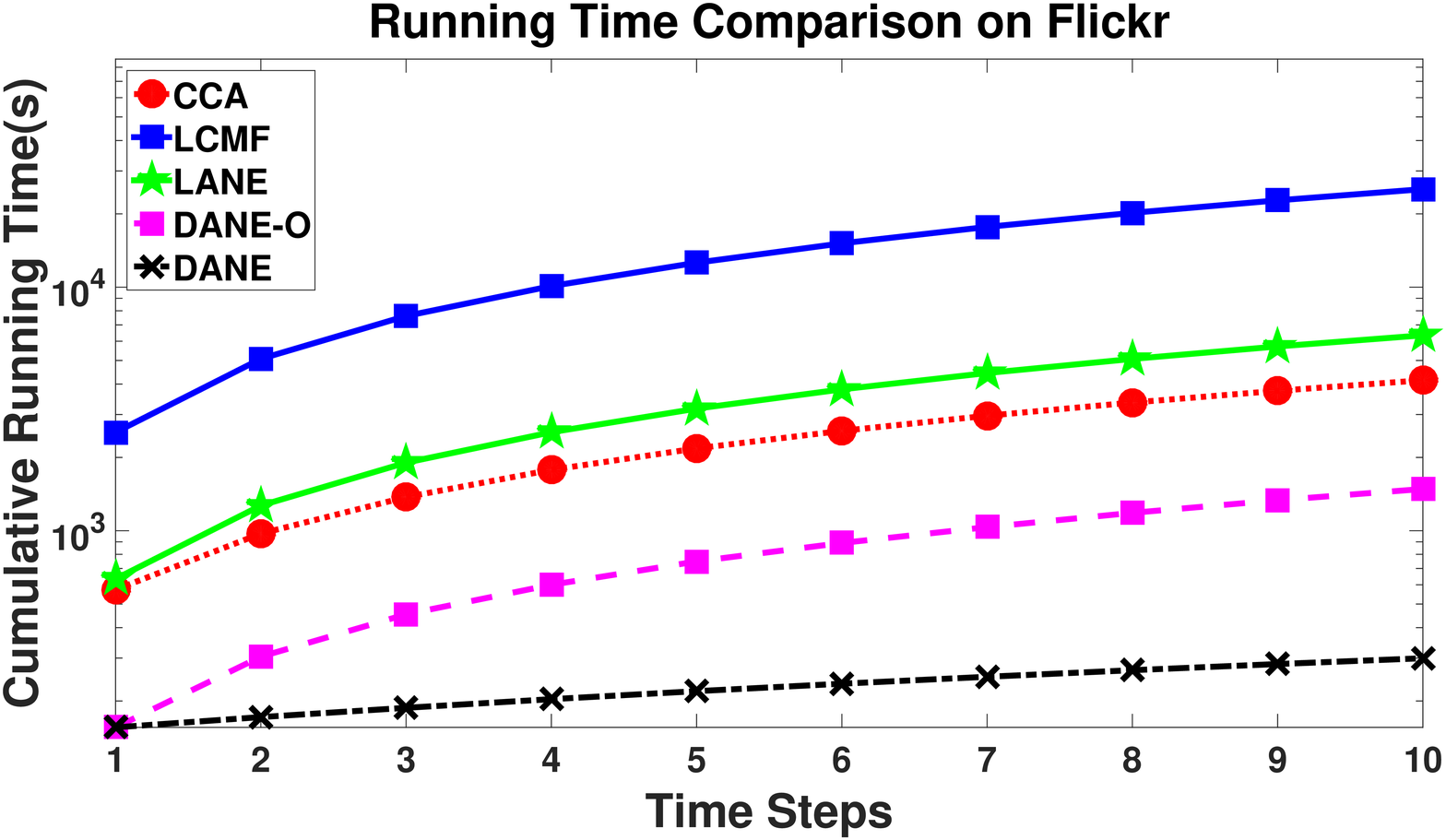}}
\end{minipage}
\begin{minipage}{0.48\textwidth}
\centering
\subfigure[Epinions\label{fig:epinionscumulative}]
{\includegraphics[width=\textwidth]{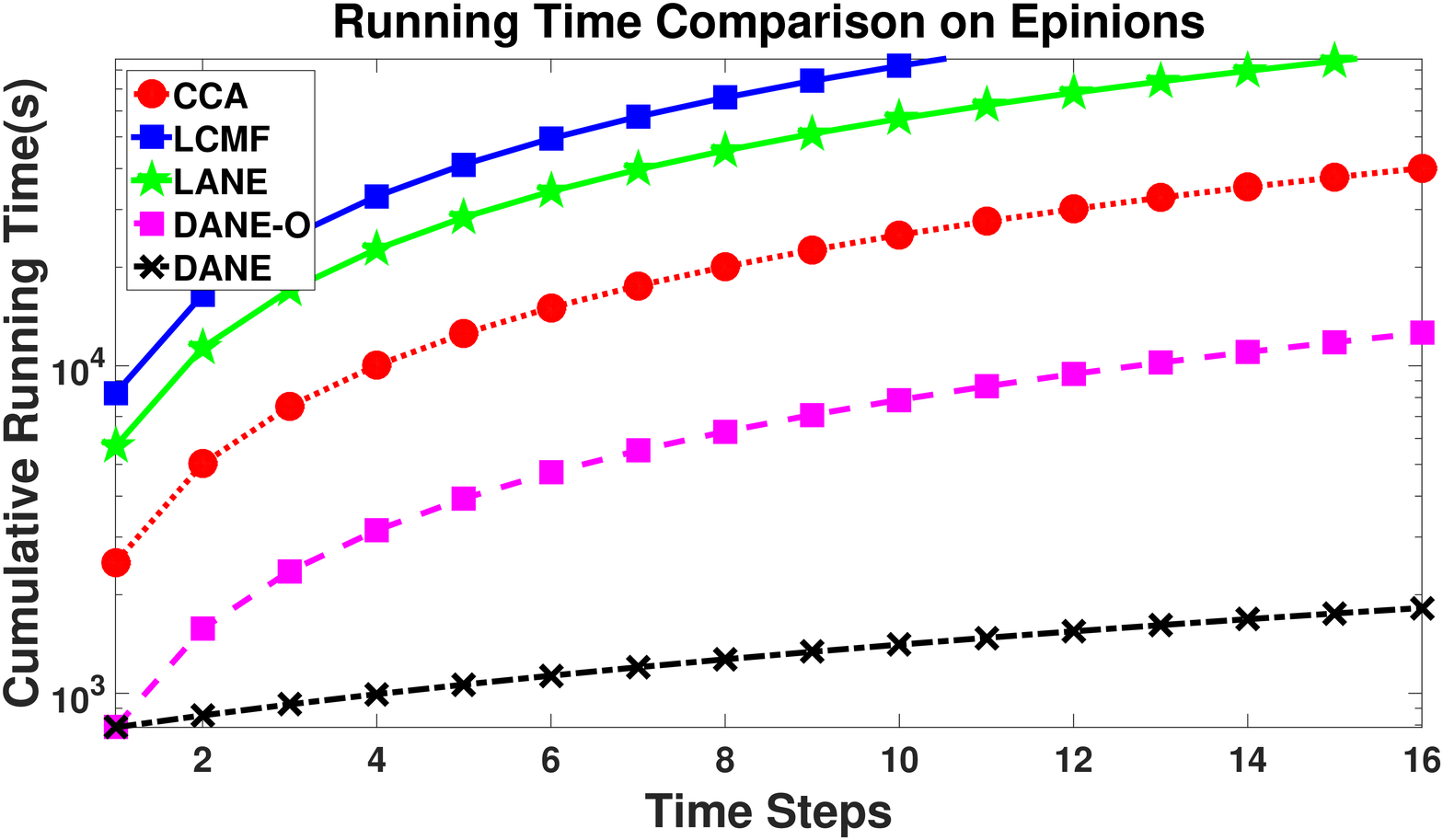}}
\end{minipage}
\begin{minipage}{0.48\textwidth}
\centering
\subfigure[DBLP\label{fig:dblpcumulative}]
{\includegraphics[width=\textwidth]{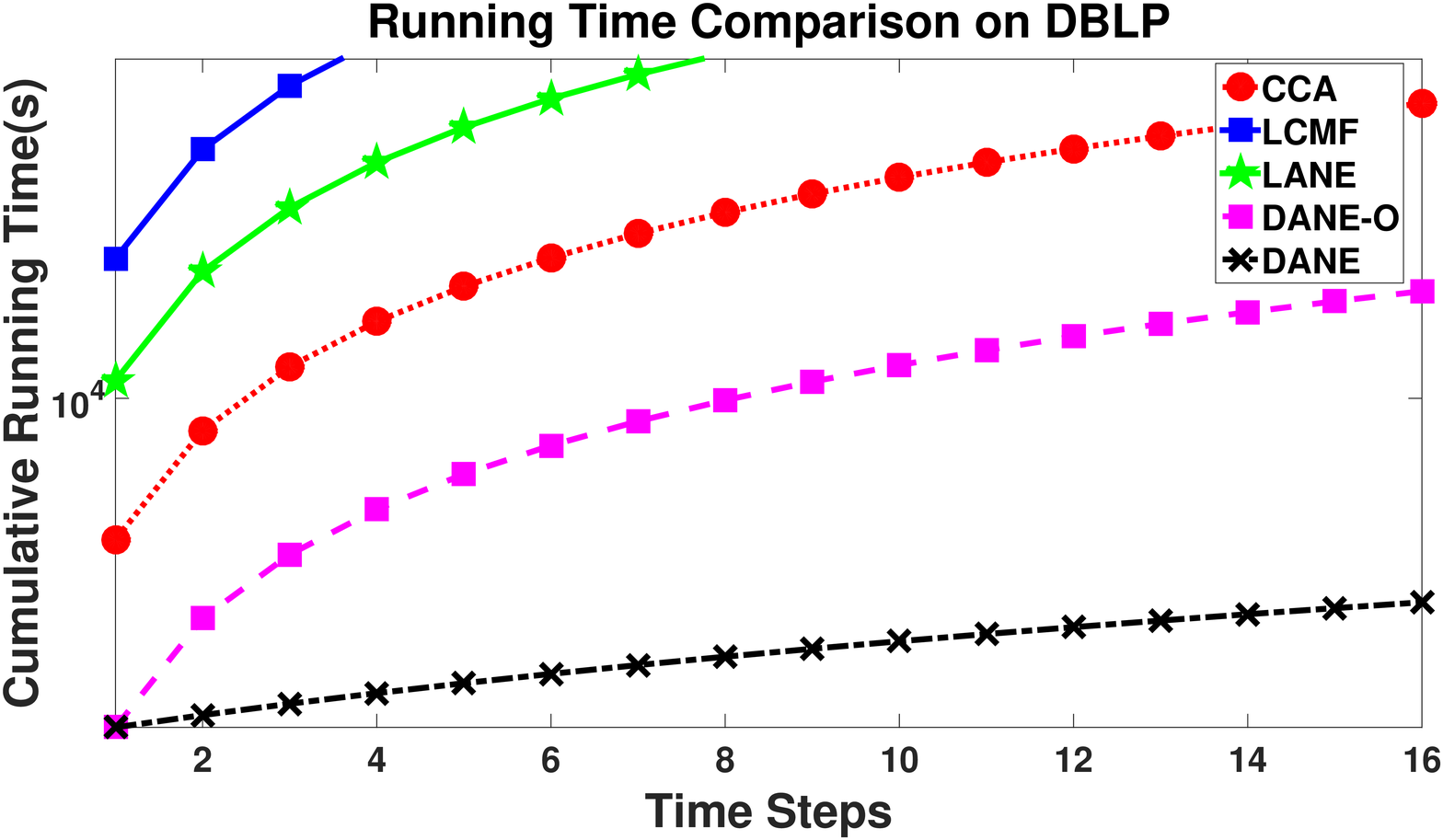}}
\end{minipage}
\vspace{-1\baselineskip}
\caption{Cumulative running time comparison.}
\label{fig:runtime}
\end{figure*}

\begin{figure*}[!t]
\centering
\begin{minipage}{0.475\textwidth}
\centering
\subfigure[BlogCatalog\label{fig:blogcatalogspeedup}]
{\includegraphics[width=\textwidth]{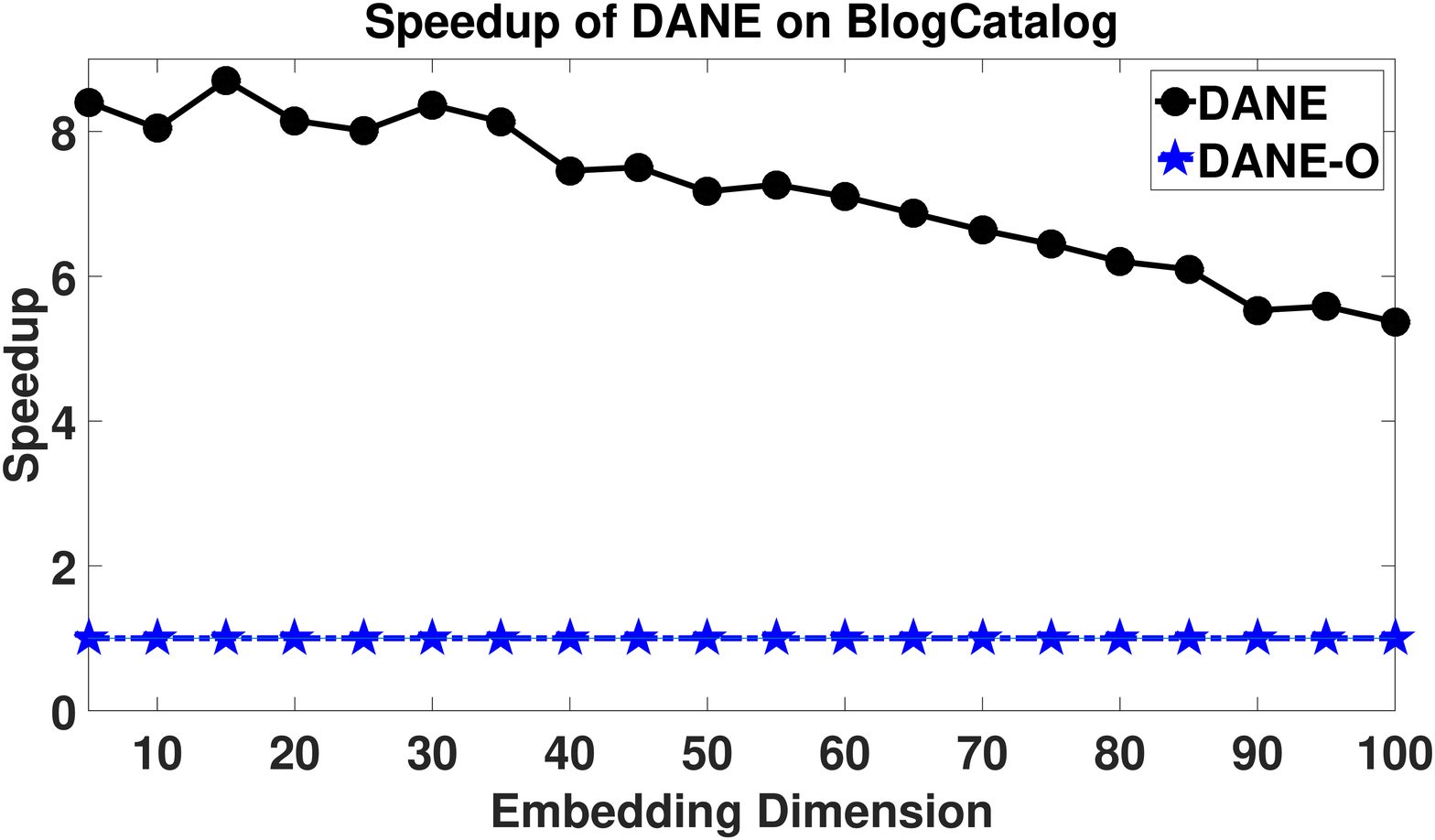}}
\end{minipage}
\begin{minipage}{0.475\textwidth}
\centering
\subfigure[Flickr\label{fig:flickrspeedup}]
{\includegraphics[width=\textwidth]{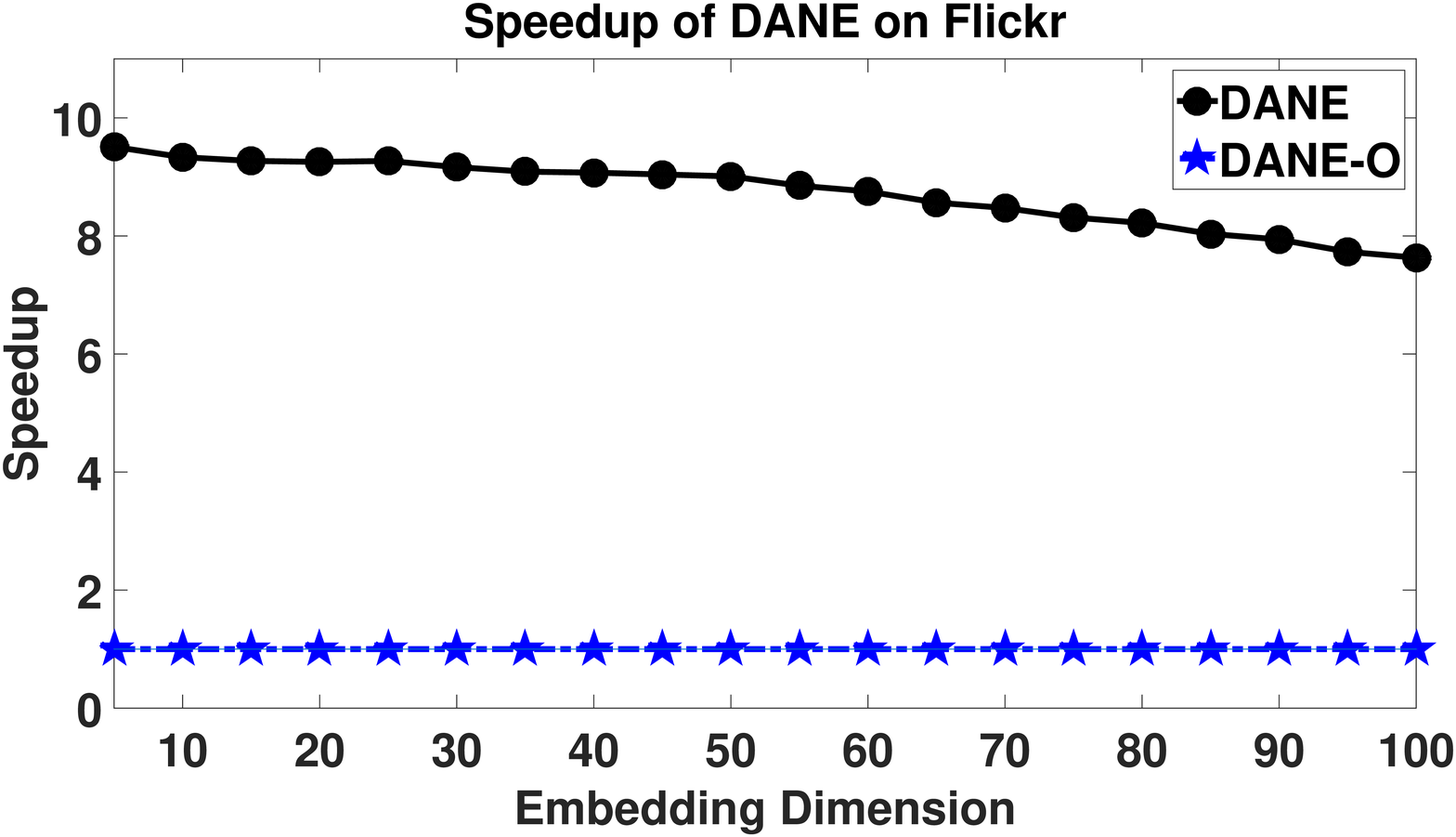}}
\end{minipage}
\begin{minipage}{0.475\textwidth}
\centering
\subfigure[Epinions\label{fig:epinionsspeedup}]
{\includegraphics[width=\textwidth]{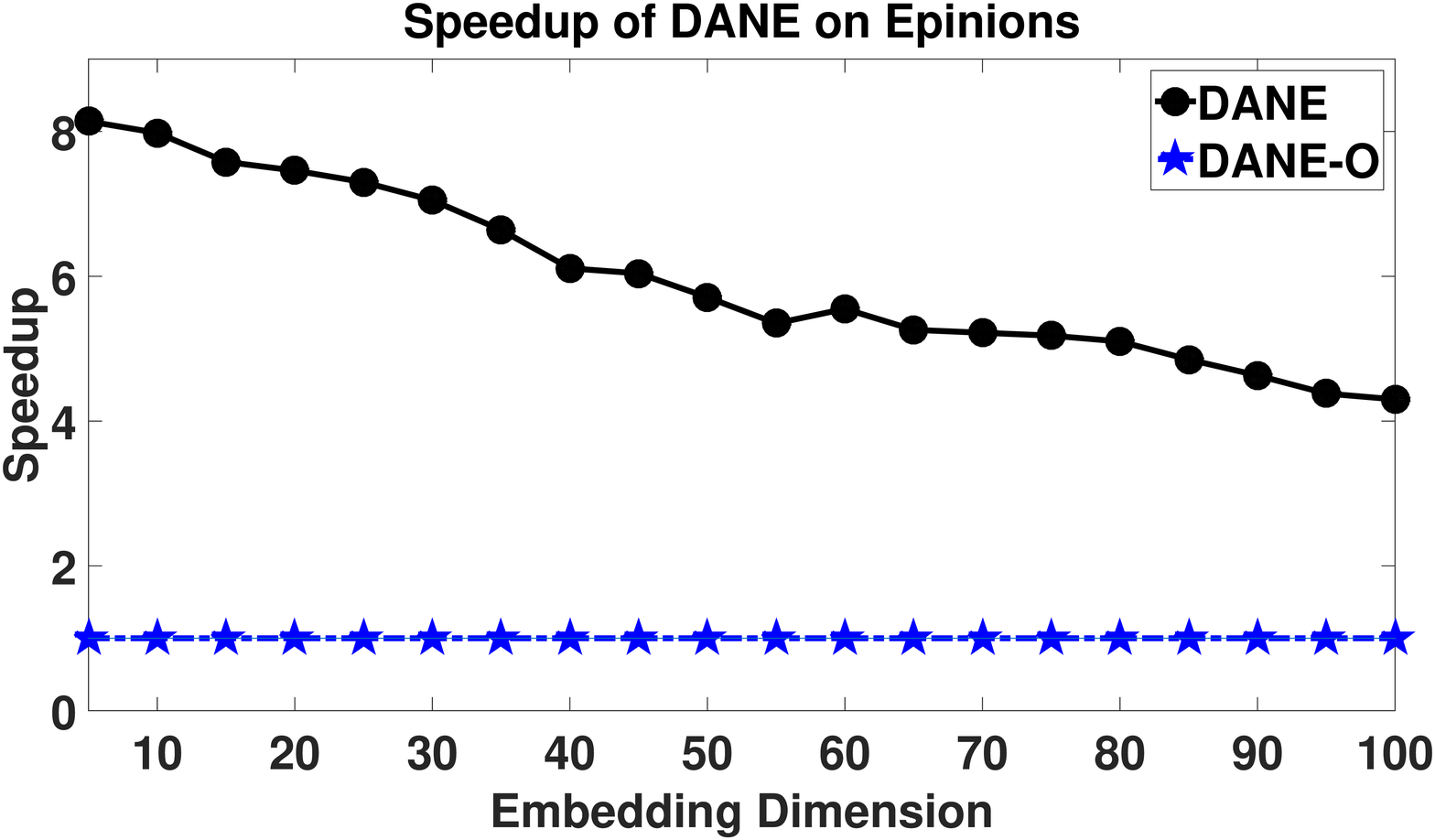}}
\end{minipage}
\begin{minipage}{0.475\textwidth}
\centering
\subfigure[DBLP\label{fig:dblpspeedup}]
{\includegraphics[width=\textwidth]{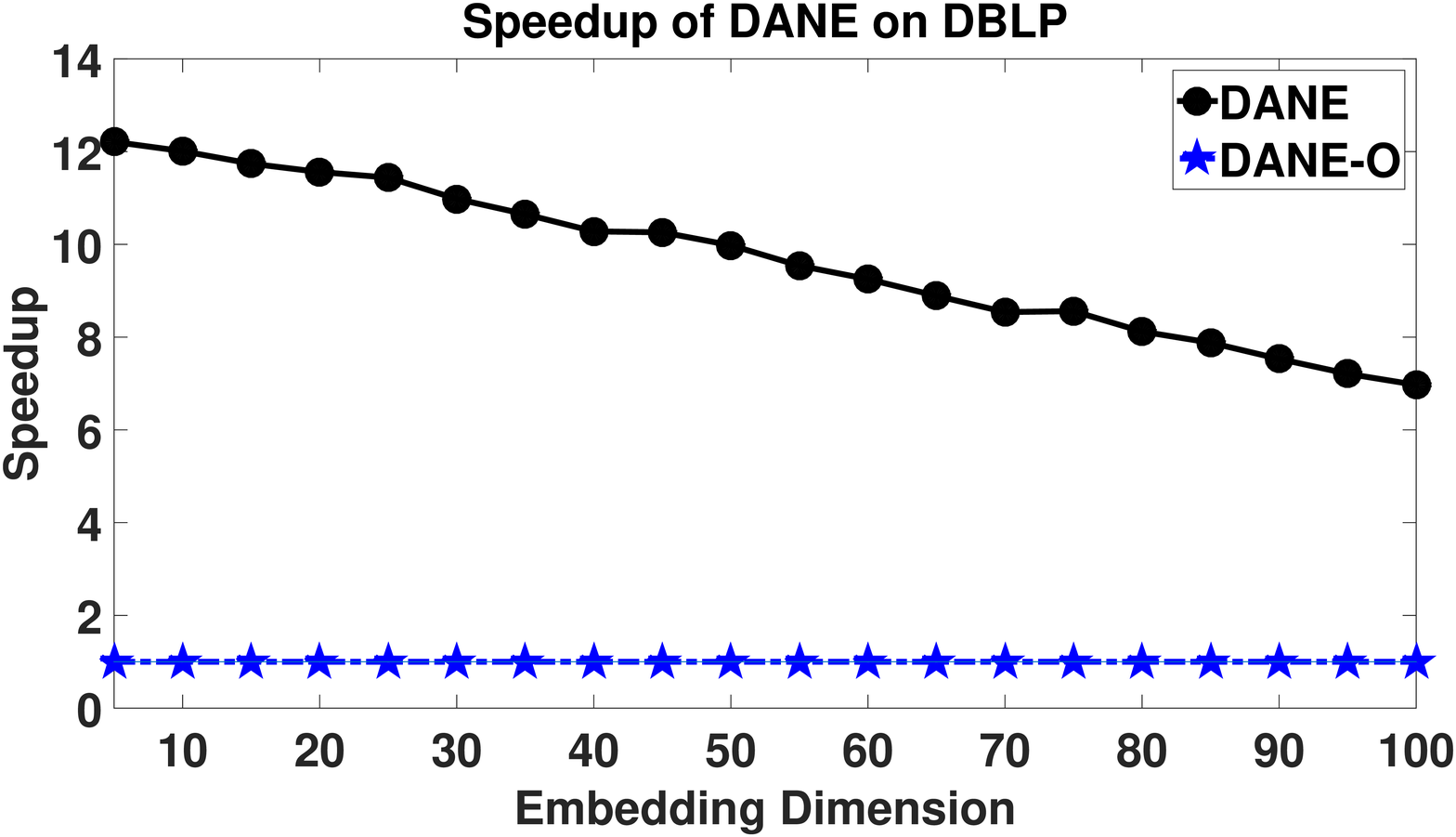}}
\end{minipage}
\vspace{-1\baselineskip}
\caption{Running time speedup of DANE against its offline version DANE-O.}
\label{fig:speedup}
\end{figure*}

To evaluate the efficiency of the proposed DANE framework, we compare DANE with several baseline methods CCA, LCMF, LANE which also use two data representations. Also, we include the offline version of DANE, i.e., DANE-O. As all these methods are not designed to handle network dynamics, we compare their cumulative running time over all time steps and plot it in a log scale. As can be observed from Figure~\ref{fig:runtime}, the proposed DANE is much faster than all these comparison methods. In all these datasets, it terminates within one hour while some offline methods need several hours or even days to run. It can also be shown that both DANE and DANE-O are much faster than all other offline methods. To be more specific, for example, DANE is 84$\times$, $21\times$ and 14$\times$ faster than LCMF, CCA and LANE respectively on Flickr dataset.
To further investigate the superiority of DANE against its offline version DANE-O, we compare the speedup rate of DANE against DANE-O w.r.t. different embedding dimensions in Figure~\ref{fig:speedup}. As can be observed, when the embedding dimension is small (around 10), DANE achieves around 8$\times$, 10$\times$, 8$\times$, 12$\times$ speedup on BlogCatalog, Flickr, Epinions, and DBLP, respectively. When the embedding dimensionality gradually increases, the speedup of DANE decreases, but it is still significantly faster than DANE-O. With all the above observations, we can draw a conclusion that the proposed DANE framework is able to learn informative embeddings for attributed networks efficiently without jeopardizing the classification and the clustering performance.

\section{Related Work}
We briefly review related work from (1) network embedding; (2) attributed network mining; and (3) dynamic network analysis.

The pioneer of network embedding can be dated back to the 2000s when many graph embedding algorithms~\cite{belkin2001laplacian,roweis2000nonlinear,tenenbaum2000global} were proposed. These methods target to build an affinity matrix that preserves the local geometry structure of the data manifold and then embed the data to a low-dimensional representation. Motivated by the graph embedding techniques, Chen et al.~\cite{chen2007directed} proposed one of the first network embedding algorithms for directed networks. They used random walk to measure the proximity structure of the directed network. Recently, network embedding techniques have received a surge of research interests in network science. Among them, Deepwalk~\cite{perozzi2014deepwalk} generalizes the word embedding and employs a truncated random walk to learn latent representations of a network. Node2vec~\cite{grover2016node2vec} further extends Deepwalk by adding the flexibility in exploring node neighborhoods. LINE~\cite{tang2015line} carefully designs an optimized objective function that preserves first-order and second-order proximities to learn network representations. GraRep~\cite{cao2015grarep} can be regarded as an extension of LINE which considers high-order information. Most recently, some deep learning based approaches are proposed to enhance the learned embeddings~\cite{wang2016structural,yang2016revisiting}.

All the above mentioned approaches, however, are limited to deal with plain networks. In many cases, we are often faced with attributed networks. Many efforts have been devoted to gain insights from attributed networks. For example, Zhu et al.~\cite{zhu2007combining} proposed a collective matrix factorization model that learns a low-dimensional latent space by both the links and node attributes. Similar matrix factorization based methods are proposed in~\cite{yang2015network,zhang2016collective}. Chang et al.~\cite{chang2015heterogeneous} used deep learning techniques to learn a joint feature representation for heterogeneous networks. Huang et al.~\cite{huang2017label} studied whether label information can help learn better feature representation in attributed networks. Instead of directly learning embeddings, another way is to perform unsupervised feature selection~\cite{tang2012unsupervised,li2016robust,cheng2017feature}. Nevertheless, all these methods can only handle static networks; it is still not clear how to learn embedding representations efficiently when attributed networks are constantly evolving over time. The problem of attributed network embedding is also related to but distinct from multi-modality or multi-view embedding~\cite{xu2013survey,kumar2011co,zhang2017react}. In attributed networks, the network structure is more than a single view of data as it encodes other types of rich information, such as connectivity, transitivity, and reciprocity.

As mentioned above, many real-world networks, especially social networks, are not static but are continuously evolving. Hence, the results of many network mining tasks will become stale and need to be updated to keep freshness. For example, Tong et al.~\cite{tong2008colibri} proposed an efficient way to sample columns and/or rows from the network adjacency matrix to achieve low-rank approximation. In~\cite{tang2008community}, the authors employed the temporal information to analyze the multi-mode network when multiple interactions are evolving. Ning et al.~\cite{ning2007incremental} proposed an incremental approach to perform spectral clustering on networks dynamically. Aggarwal and Li~\cite{aggarwal2011node} proposed a random-walk based method to perform dynamic classification in content-based networks. In~\cite{chen2015fast,chen2017eigen}, a fast eigen-tracking algorithm is proposed which is essential for many graph mining algorithms involving adjacency matrix. Li et al.~\cite{li2016toward} studied how to perform unsupervised feature selection in a dynamic and connected environment. Zhou et al.~\cite{zhou2015rare} investigated the rare category detection problem on time-evolving graphs. A more detailed review of dynamic network analysis can be referred to~\cite{aggarwal2014evolutionary}. However, all these methods are distinct from our proposed framework as we are the first to tackle the problem of attributed network embedding in a dynamic environment.

\section{Conclusions and Future Work}
The prevalence of attributed networks in many real-world applications presents new challenges for many learning problems because of its natural heterogeneity. In such networks, interactions among networked instances tend to evolve gradually, and the associated attributes also change accordingly. In this paper, we study a novel problem: how to learn embedding representations for nodes in dynamic attributed networks to enable further learning tasks. In particular, we first build an offline model for a consensus embedding presentation which could capture node proximity in terms of both network topology and node attributes. Then in order to capture the evolving nature of attributed network, we present an efficient online method to update the embeddings on the fly. Experimental results on synthetic and real dynamic attributed networks demonstrate the efficacy and efficiency of the proposed framework.

There are many future research directions. First, in this paper, we employ first-order matrix perturbation theory to update the embedding representations in an online fashion. We would like to investigate how the high-order approximations can be applied to the online embedding learning problem. Second, this paper focuses on online embedding for two different data representations; we also plan to extend the current framework to multi-mode and multi-dimensional dynamic networks.
\section*{Acknowledgements}
This material is based upon work supported by, or in part by, the National Science Foundation (NSF) grant 1614576, and the Office of Naval Research (ONR) grant N00014-16-1-2257.

\balance

\end{document}